\documentclass[pra,twocolumn,superscriptaddress,aps,showpacs,amsmath,amssymb,nofootinbib]{revtex4}
\usepackage{graphicx}
\usepackage{dcolumn}
\usepackage{bm}

\newcommand{\beq}{\begin{equation}}
\newcommand{\eeq}{\end{equation}}
\newcommand{\beqa}{\begin{eqnarray}}
\newcommand{\eeqa}{\end{eqnarray}}
\newcommand{\ket}[1]{\left| #1 \right>}

\newcommand{\expec}[1]{\big< #1 \big>}

\newcommand{\ignore}[1]{}
\newcommand{\mostra}[1]{}

\begin{document}

\title{Spin-driven spatial symmetry breaking of spinor condensates in a double-well}
\date{\today}

\author{M. Mel\'e-Messeguer}
\affiliation{Departament d'Estructura i Constituents de la Mat\`{e}ria,\\
Universitat de Barcelona, 08028 Barcelona, Spain}

\author{S. Paganelli}
\affiliation{Departament de F\'isica. Grup de F\'isica Te\`orica: Informaci\'o i Fen\`omens Qu\`antics, Universitat Aut\`onoma de Barcelona, 08193 Bellaterra (Barcelona) Spain}

\author{B. Juli\'a-D\'iaz}
\affiliation{ICFO-Institut de Ci\`encies Fot\`oniques, Parc Mediterrani de la Tecnologia, 08860 Barcelona, Spain}

\author{A. Sanpera}
\affiliation{ICREA-Instituci\'o Catalana de Recerca i Estudis Avan\c cats, Spain}
\affiliation{Departament de F\'isica. Grup de F\'isica Te\`orica: Informaci\'o i Fen\`omens Qu\`antics, Universitat Aut\`onoma de Barcelona, 08193 Bellaterra (Barcelona) Spain}

\author{A. Polls}
\affiliation{Departament d'Estructura i Constituents de la Mat\`{e}ria,\\
Universitat de Barcelona, 08028 Barcelona, Spain}

\begin{abstract}
The properties of an $F=1$ spinor Bose-Einstein condensate trapped in a double-well potential are discussed using both a mean-field two-mode approach and a simplified two-site Bose-Hubbard Hamiltonian. We focus in the region of phase space  in which spin effects lead to a symmetry breaking of the system, favoring the spatial localization of the condensate in one well. To model this transition we derive, using perturbation theory, an effective Hamiltonian that describes $N/2$ spin singlets confined in a double-well potential.
\end{abstract}
\maketitle

\section{Introduction}
\label{sec1}

For the last few years, there has been a remarkable progress in the study of 
Bose-Einstein condensates (BEC) after its first 
experimental realization in 1995 \cite{davis,anderson,bradley}. When 
the atomic confinement is achieved using magnetic traps, the 
spin degrees of freedom are frozen and the system is described by a scalar 
condensate, which is modeled by the Gross-Pitaevskii (GP) equation 
\cite{gross, pitaevskii61, dalfovo}. However, optical trapping 
\cite{stenger,stamper-kurn} has opened 
the possibility to study spinor condensates and the associated phenomena that are not present in scalar condensates \cite{ho,ohmi,zhang}.

Recently, Josephson junctions using bosonic gases in double-well traps have been investigated, and the effects of Josephson oscillations and macroscopic quantum self trapping have been experimentally observed \cite{albiez,steinhauer}.
An important feature in these systems is the appearance of strong correlations in their ground state \cite{polls10}, or the presence of a spatial symmetry breaking when the interactions are attractive and strong enough, that make the condensate to mostly localize in one well \cite{polls10b}. Strongly correlated effects in double-well potentials can be described by a simplified  Bose-Hubbard (BH) Hamiltonian \cite{fisher,demler03}. 

The aim of this work is to reveal the mechanisms that govern the spatial symmetry breaking in spinor Josephson junctions.
For this goal we derive first a mean-field two-mode approach and compare it with a fully quantum model for the two-site setup. The latter predicts also a spatial bifurcation in the condensate population that cannot be well characterized within the mean-field approach, the reason being that the bifurcation is originated by the creation and tunneling of singlets between the two wells. We focus here on the zero magnetization case, where the dimension of the Hilbert space associated to the BH Hamiltonian is maximum and the effects of the spin degrees of freedom are enhanced. 
Previous studies of Josephson junctions have focused on scalar condensates \cite{milburn,smerzi,raghavan,ananikian}, mixtures \cite{ashab, njp11}, or spinors \cite{yi,muste05,muste07,pra09}, but the effects of spin in spatial symmetric breaking have not been previously addressed.

This paper is organized as follows. We start by presenting in Sec.~II the formalism corresponding to the mean-field two-mode approach to describe a spin-1 BEC in a double-well potential. Then, we present the Bose-Hubbard model for the same setup in Sec.~III. The results obtained by the two different descriptions are 
compared in Sec.~IV.
In Sec.~V we derive a new Hamiltonian based on the tunneling of singlets that accurately accounts for the previous results, concerning the spatial symmetry breaking, and confirm the role of correlations in the spinor condensate. Finally,
the conclusions are summarized in Sec.~VI.

\section{Mean-field description of an $F=1$ spinor BEC in a double-well}
\label{sec2}

Spinor $F=1$ condensates trapped in an external potential can be described by the GP equation, which, in this case, becomes a system of three coupled non-linear equations \cite{ho}:
\beqa
i\hbar{\partial\Psi_{\pm 1}\over\partial t}&=& 
\left[{\cal H}_0 + c_2\left(n_{\pm 1} + n_0 -
    n_{\mp 1}\right) \right] \Psi_{\pm 1} + c_2 \Psi_0^2 \Psi_{\mp 1}^* ,
\nonumber \\
i\hbar{\partial\Psi_0\over\partial t}&=& 
\left[{\cal H}_0 + c_2\left(n_1 + n_{- 1}\right)
\right] \Psi_0 +2 c_2\Psi_1 \Psi_0^* \Psi_{- 1} \;,
\label{eq:gp}
\eeqa
where
\beqa
{\cal H}_0 = -{\hbar^2\over2M}\nabla^2+V_{\rm ext}+c_0 n \,.
\eeqa
For each component $\alpha=0,\pm 1$, $\Psi_\alpha(\vec{r},t)$ is the wave function of the atoms,
$n_\alpha(\vec{r},t) = \left|\Psi_\alpha(\vec{r},t)\right|^2 $ is the density of the atoms at time $t$, and $N_\alpha(t)=\int  n_\alpha(\vec{r},t)d\vec{r}$ is the number of atoms in the $\alpha$ component, which depends on time. The total density is $n(\vec{r},t) = \sum_\alpha n_\alpha(\vec{r},t)$ and the total number of particles, $N=\sum_\alpha N_\alpha(t)$, is constant.   
$V_{\rm ext}$ is the external double-well potential, which is considered symmetric. The contact interaction between atoms is characterized by the couplings $c_{0}=4\pi\hbar^2(a_0+2a_2)/3M$ and $c_2=4\pi\hbar^2(a_2-a_0)/3M$, with $M$ the mass of the atom, and $a_0$ and $a_2$ the scattering lengths describing binary elastic collisions in the channels of total spin 0 and 2, respectively.

When the condensates of the two wells are weakly linked, each of them preserve a large degree of coherence, and the two-mode ansatz \cite{smerzi} provides a good approximation of the full GP equations. This condition is fulfilled when the first two energy levels of the single particle Hamiltonian ${\cal H}_0$ are very close, forming an almost degenerate doublet, while the other ones have a much higher energy: $E_1-E_0 \ll E_2-E_0$. The low energy physics of the system can then be described using only the ground $\Phi_{\alpha +}$ and first excited $\Phi_{\alpha -}$ states of each component $\alpha$.  
In addition, as the three components have the same mass,  for $|c_2|\ll|c_0|$, one can retain only the ${\cal H}_0$ term in Eqs.~(\ref{eq:gp}), and the ground and first excited states are independent of $\alpha=0,\pm1$, namely $\Phi_+$ and $\Phi_-$. This is what is known as the single mode approximation (SMA) \cite{bigelow}. 
It is more intuitive to work with the linear combinations of these two modes, which results in one mode mostly localized in the left well $ \Phi_{L}=(\Phi_{+}+\Phi_{-}) / \sqrt{2} $, and another in the right $\Phi_{R}=(\Phi_{+}-\Phi_{-})/\sqrt{2}$. 

In the present work we also explore the properties of the condensate when $|c_2|\sim |c_0|$, and thus, out of the validity of the above approximation. However, it has been shown \cite{yi} that either when $c_2<0$ or in the particular case of $c_2>0$ and zero magnetization, the three components $\alpha=0,\pm 1$ have also the same wave function, which is now solution of the full GP equations, and the SMA can be recovered. 
The wave function of each component under the SMA, and using the two-mode ansatz is written as: 
\beqa
\Psi_{\alpha}(\vec{r},t) &=& \Psi_{\alpha L}(t) \Phi_{L} (\vec{r}) 
+ \Psi_{\alpha R}(t) \Phi_{R} (\vec{r}) \;,
\eeqa
where the time dependent coefficients are $\Psi_{\alpha j}(t)=\sqrt{N_{\alpha j}(t)} e^{i\phi_{\alpha j}(t)}  $. The number of particles at each side, neglecting the small overlap between the left and right modes, is 
$N_{\alpha L} = \int_{-\infty}^0 dx \int_{-\infty}^{+\infty} dy \int_{-\infty}^{+\infty} dz \left|\Psi_{\alpha} (\vec{r},t) \right|^2 $ and $N_{\alpha R}= 
\int^{+\infty}_0 dx \int_{-\infty}^{+\infty} dy \int_{-\infty}^{+\infty} dz
\left|\Psi_{\alpha} (\vec{r},t) \right|^2 $.

We introduce this two-mode ansatz into the GP equation (\ref{eq:gp}) to obtain the two-mode equations for the spinor condensate \cite{njp11,pra09} (see Appendix~\ref{ap1}). These equations result in a system of eight coupled non-linear differential equations relating the population imbalance of each component, $z_\alpha(t) = \big(N_{\alpha L}(t)-N_{\alpha R}(t)\big)/N_{\alpha}(t)$, the phase difference $\delta\phi_{\alpha}(t) = \phi_{\alpha R}(t) - \phi_{\alpha L}(t)$, the population of the $m=0$ component, $N_0(t)$, and the phase $\Delta\phi_L(t)=2\phi_{0 L}(t) - \phi_{-1 L}(t)-\phi_{+1 L}(t)$. The parameters that control the dynamics are the tunneling rate $K$ and the interaction energies $\xi_{(0,2)}$:
\beqa
K\quad  &=& -\int d^3r \bigg[{\hbar^2\over 2M} \nabla \Phi_L \cdot \nabla \Phi_R +
\Phi_L V_{\rm ext} \Phi_R \bigg]
\nonumber \\
\xi_{0(2)} &=& {c_{0(2)}\over 2} \int d^3r \Phi_{L}^4(\vec{r}) \;.
\eeqa

\section{Bose-Hubbard model for $F=1$ spinors in a double-well}
\label{sec3}

The generalization of the two-site BH Hamiltonian for an $F=1$ BEC is
\cite{demler}:
\beqa
H&=&
- J \sum_{\alpha=0,\pm 1} 
\Big(\hat{a}_{\alpha L}^\dagger \hat{a}_{\alpha R} +
\hat{a}_{\alpha R}^\dagger \hat{a}_{\alpha L}  \Big)
\nonumber \\
&& +{U_0\over 2} \Big(\hat{N}_L(\hat{N}_L-1) + \hat{N}_R(\hat{N}_R-1)\Big)
\label{eq:bh} \\
&& +{U_2\over 2} \Big(\hat{\bf S}_L^2 - 2\hat{N}_L + \hat{\bf S}_R^2 - 2\hat{N}_R \Big) +\sum_{j=L,R} \varepsilon_j\hat{N}_j 
 \;,
\nonumber
\eeqa
where $J$ is the tunneling coupling between sites, $U_0$ is equivalent to scalar interactions, and therefore spin-independent, while $U_2$ derives from the spin interactions. The operator $\hat{a}_{\alpha j}\,\big(\hat{a}_{\alpha j}^\dagger\big)$ is the annihilation (creation) operator of a particle of component $m_\alpha$ in the $j$-th site, and obeys the usual bosonic commutation rules, $[\hat{a}_{\alpha j},\hat{a}_{\beta k}] = [\hat{a}_{\alpha j}^\dagger,\hat{a}_{\beta k}^\dagger] = 0 $ and $[\hat{a}_{\alpha j},\hat{a}_{\beta k}^\dagger] = \delta_{\alpha \beta}\delta_{j k}$. The number of particles populating the $\alpha$ component is defined as $\hat{N}_{\alpha j} = \hat{a}_{\alpha j}^\dagger \hat{a}_{\alpha j}$, and the total number of particles in the $j$-th site is $\hat{N}_j = \sum_\alpha \hat{N}_{\alpha j}$. The operator $\hat{ \bf S}_j$ is a pseudo-angular momentum operator in the $j$-site defined as:
\beqa
\hat{S}_j^{(z)} &=& \hat{N}_{+1 j} - \hat{N}_{-1 j} = \hat{a}_{+1 j}^\dagger
\hat{a}_{+1 j} - \hat{a}_{-1 j}^\dagger \hat{a}_{-1 j}
\nonumber \\
\hat{S}_j^{(+)} &=& \sqrt{2}\Big(
\hat{a}_{+1 j}^\dagger \hat{a}_{0 j} + \hat{a}_{0 j}^\dagger \hat{a}_{-1 j}
\Big) 
\nonumber \\
\hat{S}_j^{(-)} &=& \hat{S}_j^{(+)\dagger} \; ,
\label{eq:s}
\eeqa
with $\big[ \hat{S}_j^{(+)} , \hat{S}_k^{(-)} \big] = 2\delta_{jk} \hat{S}_j^{(z)} $ and $ \big[ \hat{S}_j^{(z)} , \hat{S}_k^{(\pm)} \big] = \pm\delta_{jk} \hat{S}_j^{(\pm)} $. Finally, $\varepsilon_j$ acts as a bias that breaks the degeneracy between wells and controls the spatial symmetry breaking. 

For convenience, we introduce the Fock basis, that is labeled by the number of particles of each component in each well: $\big\{\ket{N_{-1L}, N_{-1R}, N_{0L}, N_{0R}, N_{+1L}, N_{+1R}}\big\}$, with a fixed total number of particles, $N = \sum_{\alpha j} N_{\alpha j}$, and magnetization, $M =\sum_j (N_{+1 j}-N_{-1 j})$. The minimum dimension of the Hilbert space spanned by this basis is $N+1$ and corresponds to maximum magnetization ($M=N$ or $M=-N$). In this case all the particles are in the same state $m=+1$ or $m=-1$, and the system reduces to the single component case with an effective interaction $U_0+U_2$. When the magnetization decreases, the dimension grows and reaches its maximum $(N+2)(N+4)(12+6N+N^2)/96$ for $M=0$, growing with $N$ as $N^4$. 

For our subsequent discussion, it is useful to introduce another basis, which is defined as the simultaneous
eigenstates of the number of particles $\hat{N}_{j}$, the angular momentum $\mathbf{\hat{S}}_{j}^{2}$, and the magnetization $\hat{S}^{(z)}_{j}$ in each $j=L,R$:
\begin{eqnarray}\label{eqn:base2}
\hat{N}_{j}\left|s_{j},m_{j},n_{j}\right\rangle  & = & n_{j}\left|s_{j},m_{j},n_{j}\right\rangle , \nonumber \\
\mathbf{\hat{S}}_{j}^{2}\left|s_{j},m_{j},n_{j}\right\rangle  & = & s_{j}(s_{j}+1)\left|s_{j},m_{j},n_{j}\right\rangle , \nonumber \\
\hat{S}^{(z)}_{j}\left|s_{j},m_{j},n_{j}\right\rangle  & = & m_{j}\left|s_{j},m_{j},n_{j}\right\rangle,
\end{eqnarray}
and where the sum $s_{j}+n_{j}$ has to be even \cite{wu}.

It is interesting to compare the results obtained with the two descriptions, described in Sects.II and III. 
The standard procedure consists in replacing the field operators $\hat{a}_{\alpha j}\big(\hat{a}_{\alpha j}^\dagger \big)$ by c-numbers $\sqrt{N_{\alpha j}} {\rm e}^{i\phi_{\alpha j}}\big(\sqrt{N_{\alpha j}} {\rm e}^{-i\phi_{\alpha j}}\big)$, to obtain a semiclassical Hamiltonian $H_{\rm s}$ (see Appendix \ref{ap2}). Assuming that the variables $(N_{j\alpha}, \phi_{j\alpha})$ are canonical conjugate, we obtain the equations of motion using Hamilton's equations $\dot{N}_{j,\alpha} = \partial H_{s} / \partial \phi_{j,\alpha}$ and $\dot{\phi}_{j,\alpha} = -\partial H_{s} / \partial N_{j,\alpha}$.

Remarkably, the dynamics predicted by these equations and by the mean-field two-mode equations derived in Sec.~\ref{sec2} is exactly the same, when $J=K$, $U_0 = 2 \xi_0$ and $U_2 = 2\xi_2$. Moreover, it is also in agreement with the results reported in Ref.~\cite{pra09}, which were obtained in the limit $z_\alpha \sim 0$, $\delta\phi_\alpha \sim 0$, $\Delta\phi_L\sim 0$ and $M=0$.

In the case of only one component, a similar result was derived when the mean-field two-mode approximation was compared with the semiclassical version of the two-site BH Hamiltonian \cite{ananikian}.

\section{Ground state properties}
\label{sec4}

In this section, first we review the results for the ground state of a spinor $F=1$ condensate confined in a single well, and then discuss the results found with the double-well. 

\subsection{Single well}
\label{ssec4a}

\subsubsection{Mean-field description}
\label{sssec4a1}

The Gross-Pitaevskii equations, Eqs.~(\ref{eq:gp}), are invariant under the gauge transformation $\Psi\to {\rm e}^{i\theta}\Psi $, where $\Psi=(\Psi_{-1},\Psi_0,\Psi_{+1})$, and any spin rotation $\Psi\to {\cal U}(\alpha,\beta,\tau)\Psi$, where ${\cal U}(\alpha,\beta,\tau) = {\rm e}^{-iF_z\alpha}{\rm e}^{-iF_y\beta} {\rm e}^{-iF_z\tau}$. $F_i$ are the corresponding spin-1 matrices, and $(\alpha,\beta,\tau)$  the Euler angles \cite{ho}, that define the spin rotation, with ranges $\theta,\alpha,\tau\in(-\pi,\pi)$ and $\beta\in(-\pi/2,\pi/2)$. This invariance produces a degeneracy in the eigenstates of the Hamiltonian \cite{ho}. 

For the polar case, $c_2>0$, the degenerate ground state is:
\beqa
\ket{\Psi_{\rm g.s.}}_{c_2>0} = {\rm e}^{i\theta} \left(\begin{array}{c}
-{1\over \sqrt{2}} {\rm e}^{-i\alpha} \sin\beta \\
\cos\beta \\
{1\over \sqrt{2}} {\rm e}^{i\alpha} \sin\beta 
\end{array} \right) \;, \label{eq:mf-u2p}
\eeqa
that have an average number of atoms in the different components of 
$(N_{-1},N_{0},N_{+1}) =$ $(\sin^2\beta$, $2\cos^2\beta$, $\sin^2\beta)/2$, depending only on $\beta$.

For the ferromagnetic case, $c_2<0$, the ground state set is:
\beqa
\ket{\Psi_{\rm g.s.}}_{c_2<0} = {\rm e}^{i\theta-\tau} \left(\begin{array}{c}
{\rm e}^{-i\alpha} \cos^2{\beta\over 2} \\
\sqrt{2} \cos{\beta\over 2}\sin{\beta\over 2} \\
{\rm e}^{i\alpha} \sin^2{\beta\over 2}
\end{array} \right) \;, \label{eq:mf-u2n}
\eeqa
with $(N_{-1},N_{0},N_{+1}) =$ $(\cos^4(\beta/2)$, $(\sin^2\beta)/2$, $\sin^4(\beta/2))$.

Note that, for the particular case of $M=0$, the angle $\beta$ can take any value for $c_2>0$ whereas only one value is allowed for $c_2<0$, i.e. $\beta=\pi/2$.

\subsubsection{Quantized description}

As the number of particles is fixed, in the quantized Hamiltonian:
\beqa
H = {U_0\over 2}\hat{N} (\hat{N}- 1) + { U_2\over 2}  (\hat{{\bf S}}^2 - 2\hat{N}) \; , 
\eeqa
the only relevant term to find the ground state is the one proportional to $U_2$ \cite{bigelow}.  In the following, we consider an even number of particles, although similar arguments apply for an odd $N$. 

For $U_2>0$, the ground state has the minimal value of $\hat{S}^2$, i.e. $s=0$, and there is only one possible magnetization $M=0$. In the basis labeled by $\ket{N_{-1}, N_0, N_{+1}}$, the ground state can be written as \cite{bigelow}:
\beqa 
\ket{\Psi_{\rm g.s.}} &=& \sum_{k=0}^{N/2} A_k \ket{k,N-2k,k} \;,  \label{eq:par1} \\
A_k &=& -\sqrt{N-2k+2\over N-2k+1} A_{k-1} \;,
\eeqa 
which gives an average number of atoms of $\expec{\hat{N}_{+1}}=\expec{\hat{N}_{0}}=\expec{\hat{N}_{-1}}=N/3$ and large fluctuations in each component, e.g. $\expec{\Delta \hat{N}_0} \approx 2N/\sqrt{5}$ for $N\gg 1$. 

On the other hand, when $U_2<0$ the ground state maximizes the pseudo-spin, so that $s=N$, and the magnetization can take any even value from $M=0$ to $M=\pm N$. These states have the general form:
\beqa
\ket{\Psi_{\rm g.s.}} &=& \sum_k B_k^{(M)} \ket{k, N-2k - M, k+M} \label{eq:par2} \;,
\eeqa
and the values of $B_k^{(M)}$ are determined starting from the state
$\ket{N,0,0}$, which has $M=-N$ and only $B_N^{(-N)}=1$, and applying repeatedly the rising operator $S^{(+)}$. In this case, the amplitudes $B_k^{(M)}$ have a narrow distribution around a certain $k$ value, which indicates that the number of particles in each component is reasonably well defined \cite{bigelow}. 

Finally, note that in both cases the BH description is compatible with the mean-field results presented in Sec.~\ref{sssec4a1} when $N\gg 1$.

\subsection{Double-well}\label{sec:double}

\subsubsection{Mean-field description}

To fix ideas and notations, let us first review the  results of the mean-field two-mode approximation \cite{smerzi} for a scalar condensate. The Hamiltonian in this case is very simple:
\beqa
H_{\rm s.c.} = -\sqrt{1-z^2} \cos\delta\phi +  \Lambda z^2 \; ,
\eeqa
where $z$ is the population imbalance, $\delta\phi$ the phase difference between wells and $\Lambda= NU_0/(2J)$ is the only free parameter of the system, with $N$ the number of particles, $U_0$ is proportional to the atom-atom scattering length and $J$ the tunneling rate. The ground state can be found minimizing the Hamiltonian $H_{\rm s.c.}$, and has 
$\delta\phi=0$ and an imbalance that depends on $\Lambda$: $z=0$ when $\Lambda \geq - 1$ and $z=\pm \sqrt{1-1/\Lambda^2}$ when $\Lambda < -1$. This means that for the critical value $\Lambda_c = -1$, the population imbalance of the ground state bifurcates into two different degenerate solutions, each one corresponding to the atomic cloud mostly localized in a different well. 

\begin{figure}
\includegraphics[width=0.85\columnwidth,angle=0, clip=true]{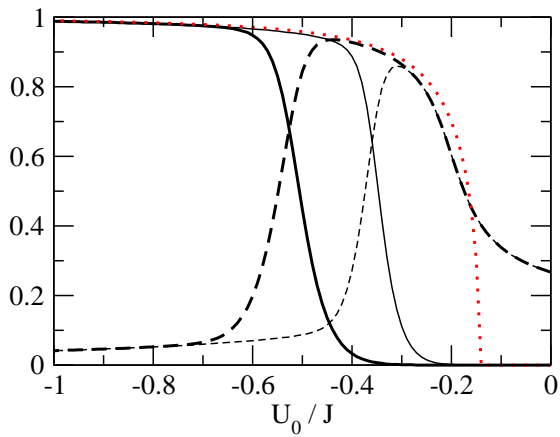}
\caption{In dotted-red we plot the population imbalance $z$ of the ground state of a scalar condensate obtained with the mean-field two-mode description, with $N=14$. The black lines correspond to the two-site BH predictions for a scalar condensate with $N=14$, and correspond to the population imbalance $z$ (solid), and its dispersion $\sigma_z$ (dashed) for two different values of $\varepsilon_L/J=10^{-4}$ (thin) and $\varepsilon_L/J=10^{-6}$ (thick). In both cases $\varepsilon_R/J=0$.   
\label{fig-1}}
\end{figure}

In Fig.~\ref{fig-1} we plot the population imbalance of the ground state as a function of $U_0/J$. For weak interactions, $z=0$ and the condensate has the same amount of atoms in each well. At the bifurcation point, defined by $\Lambda=\Lambda_c=-1$, the interaction for $N=14$ is $U_0^c/J = -2\Lambda_c/N = -0.143$, and the population imbalance bifurcates into two different non-zero solutions $\pm\sqrt{1-1/\Lambda^2}$. However, only the positive branch, which corresponds to having more atoms in the left well, is plotted in Fig.\ref{fig-1}.

For a spinor $F=1$ condensate with zero magnetization, we obtain the ground state by minimizing the semiclassical two-mode Hamiltonian. We assume that the ground state has the same population imbalance for each component $\alpha=0,\pm 1$, $z_\alpha \equiv z$, and the same phase difference $\delta\phi_\alpha \equiv \delta\phi$. 

\begin{figure}
\includegraphics[width=0.85\columnwidth,angle=0, clip=true]{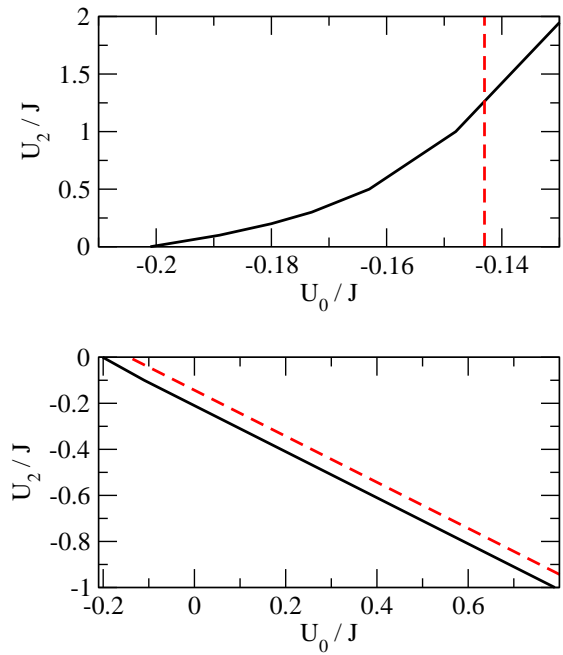}
\caption{Characterization of the bifurcation for $U_2/J>0$ (top) and $U_2/<0$ (bottom) for $N=14$ and $M=0$. In dashed-red we plot the bifurcation point predicted by the mean-field two-mode description, that corresponds to $\Lambda_c=-1$. And in solid-black the bifurcation obtained with the two-site Bose-Hubbard model, which corresponds to the value of $U_0/J$, for each $U_2/J$, where the dispersion $\sigma_z$ has an inflexion point. 
\label{fig-2}}
\end{figure}

The solution for $z$, $\delta\phi$, is the same as in the scalar case but with an interaction parameter  $\Lambda=NU_0/(2J)$ for $U_2>0$, and $\Lambda=N(U_0+U_2)/(2J)$ for $U_2<0$. In Fig.~\ref{fig-2} we plot the bifurcation point, defined by $\Lambda=\Lambda_c=-1$, as a function of both $U_0/J$ and $U_2/J$. 

The distribution in the number of particles and the phase $\Delta\phi_L$ for the ground state are found to be $(N/2,0,N/2)$ and $\Delta\phi_L=\pi$ for $U_2>0$, and $(N/4,N/2,N/4)$  and $\Delta\phi_L=0$ for $U_2<0$. These solutions are compatible with the semiclassical results for the single well. 

\subsubsection{Two-site Bose-Hubbard} \label{sec:double-twosite}

\begin{figure}[t]
\includegraphics[width=0.85\columnwidth,angle=0, clip=true]
{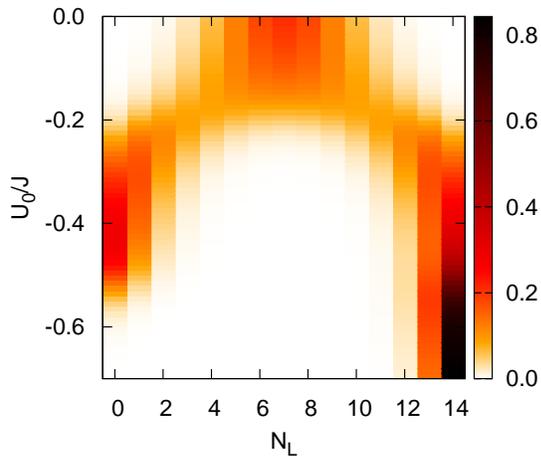}
\caption{Representation of the ground state of a scalar condensate with $N=14$, $\varepsilon_L/J=10^{-6}$ and $\varepsilon_R/J=0$. The color corresponds to the value of the coefficients $|c_{N_L}|^2$, which are plotted as a function of the number of particles in the left well $N_L$ and the interaction $U_0/J$. 
\label{fig-3}}
\end{figure}

Again for clarity, we first review the results of a scalar condensate, described by the two-site Bose-Hubbard Hamiltonian:
\beqa
H&=& \sum_{j=L,R} \varepsilon_j\hat{N}_j 
- J \Big(\hat{a}_{L}^\dagger \hat{a}_{R} +
\hat{a}_{R}^\dagger \hat{a}_{ L}  \Big)
\nonumber \\
&& +{U_0\over 2} \Big(\hat{N}_L(\hat{N}_L-1) + \hat{N}_R(\hat{N}_R-1)\Big) \,.
\label{eq:sc-bh}
\eeqa
We use the Fock basis, labeled by $\ket{N_L,N-N_L}\equiv\ket{N_L}$, and diagonalize this Hamiltonian to find the ground state, $\ket{gs} = \sum_{N_L} c_{N_L} \ket{N_L}$. The coefficients $|c_{N_L}|^2$ of the many-body state are plotted in Fig.\ref{fig-3} as a function of the interaction $U_0/J$. For weak interactions, the distribution of $|c_{N_L}|^2$ is peaked around the Fock state $\ket{N/2,N/2}$. For stronger interactions the ground state becomes strongly correlated and has two different peaks, each of them centered around an imbalanced Fock state. Finally, the ground state peaks towards large $N_L$, meaning that on average the atoms are more likely to be found in the left well than in the right well. As the interaction is increased, the distribution of $|c_{N_L}|^2$ is peaked around a Fock state closer to the state $\ket{N,0}$ \cite{polls10}.
It is worth noting that in the quantum case, localization only appears in the thermodynamic limit, when the ground and the first excited states are degenerate and a spontaneous symmetry breaking occurs. However, when the energy spacing between the ground and first excited states becomes smaller than the precision of the numerical calculations, the system behaves as if it had an effective degeneration, and asymmetric states can be achieved. For a fixed number of particles, this phenomenon only occurs when $U_0$ is negative and sufficiently large.

To drive the localization on the left well we introduce a small bias $\varepsilon_L$ and $\varepsilon_R=0$. The effect of this bias in the localization can be seen in Fig.~\ref{fig-1}, where we plot the expected value of the population imbalance operator, $\hat{z} = (\hat{N}_L-\hat{N}_R)/N$ and its dispersion $\sigma_z=\sqrt{<\hat{z}^2>-<\hat{z}>^2}$, for the ground state for two different values of $\varepsilon_L$. One can see that the population imbalance $z$ depends on the bias, and that the behavior of $z$ and $\sigma_z$ before the symmetry breaking is independent of the bias \cite{polls10b}. 

This allows us to define a quantum analog of the semiclassical bifurcation point, independent of the bias,  as the value of $U_0/J$ where the dispersion $\sigma_z$ has an inflexion point. In Fig.~\ref{fig-1} we can see that this point corresponds to $U_0/J\sim -0.2$, and in Fig.\ref{fig-3} we can see that at this value of the interaction the many-body ground state is very broad and approximately goes from having one to two peaks. See Appendix~\ref{ap3} for an estimation of the value of $U_0$ corresponding to the bifurcation point as a function of $J$ and $N$.

For the spin $F=1$ condensate, the ground state is found by diagonalizing the BH Hamiltonian Eq.~(\ref{eq:bh}) for a fixed number of particles and magnetization. The distribution of the number of particles of this state turns out to be only dependent of the sign of $U_2$ and equal to the distribution found for the single well, described by Eqs.~(\ref{eq:par1}) and (\ref{eq:par2}). This is because neither $J$ nor $U_0$ depend on the spin component $m$, and only the $U_2$ term determines the population of the components. Therefore, there are only two relevant parameters to characterize the GS: the total population imbalance $\hat{z} = {1\over N} \sum_\alpha \big(\hat{N}_{\alpha L} -\hat{N}_{\alpha R}\big)$ and its dispersion $\sigma_z= \sqrt{<\hat{z}^2> - <\hat{z}>^2}$. 
In the following, we focus in the case of $M=0$. 

Applying the same arguments used for the single well, for $U_2>0$ the ground state minimizes the pseudo-spin in each side, $s_L=s_R=0$, so that effectively the $U_2$ term in the Hamiltonian for the ground state reduces to the constant term $-U_2 N$. The Hamiltonian is equivalent to a scalar Hamiltonian Eq.~(\ref{eq:sc-bh}) with interaction $U_0$, and thus, the bifurcation is independent of $U_2$. 

When $U_2<0$ the ground state maximizes the pseudo-spin in both sides, so $s_L=N_L$ and $s_R=N_R$, and the spin-changing term of the Hamiltonian reduces to:
\beqa
{U_2\over 2} \bigg[\hat{N}_L(\hat{N}_L-1) + \hat{N}_R(\hat{N}_R-1) \bigg] \;.
\eeqa
This allows us to consider the Hamiltonian as an scalar one, Eq.~(\ref{eq:sc-bh}), with an effective interaction $U_0+U_2$.

The quantum analog of the semiclassical bifurcation is defined in a similar way as in the scalar case, and is taken, for every $U_2/J$,  as the value of $U_0/J$ for which $\sigma_z$ has an inflexion point. In Fig.~\ref{fig-2} we plot this point for different values of $U_2$, were we can see that for $U_2>0$ (top) this point depends on the strength of the interaction in contrast to the mean-field two-mode predictions, also plotted. This means that the many-body state delocalizes when the value of $U_2$ is increased, and at some point, and due to the bias, localizes in the left region of the Fock space. This discrepancy between the full quantum and the semiclassical two-mode descriptions will be explained in the next section. On the other hand, when $U_2<0$, the bifurcation point has exactly the same dependence with the strength of $U_2$ as the mean-field two-mode prediction.

\section{Spin driven symmetry breaking}
\label{sec5}

To understand why the system delocalizes as we increase the value of 
$U_2$, we study the composition of the ground state of the system. 
We characterize the \emph{seniority} of the ground state \cite{senio}, 
as the number of pairs of atoms that are coupled to total spin 
0 in the many-body state. As we will see in the following, the symmetry 
breaking described in the previous section is directly linked to the 
presence of a large amount of spin-zero pairs in the many-body ground state.

It is useful to define the creation operator of a spin singlet:
\beqa
\hat{\Theta}^{\dagger}=\hat{a}_{0}^{\dagger2}-2\hat{a}_{1}^{\dagger}\hat{a}_{-1}^{\dagger} \,,
\eeqa
which creates a two-particle spin-zero state. It can be applied to
the vacuum $k$ times to produce $k$ singlets. This state, in the basis defined by (\ref{eqn:base2}) is:
\beqa \label{eqn:singletstate}
\left|0,0,2k\right\rangle_j =\frac{\left(\hat{\Theta}^{\dagger}\right)^{k}}{y(1)\ldots y(k)}\left|0,0,0\right\rangle_j \,,
\eeqa
with $y(k)=\sqrt{2k\left(2k+1\right)}$ and $j=L,R$.

Fixing the total number of particles $N$ and the total magnetization
$M$, the basis can be labeled only by four quantum numbers:
\beqa
\left|s_{L},m_{L},n_{L}\right\rangle \left|s_{R},M-m_{L},N-n_{L}\right\rangle =\left|s_{L},s_{R},m_{L},n_{L}\right\rangle. \nonumber \\
\eeqa

In Fig.~\ref{fig-2} one can see that the spin interaction influences the occurrence of the bifurcation 
and its behavior changes depending on the sign of $U_2$. 
The case of $U_{2}<0$ is easily understood, since in this regime the spin on each site tends to be as large as 
possible. When the system starts to localize one can assume that $\bold{S}^2_j \simeq n_j(n_j+1)$, so that
Eq.~(\ref{eq:bh}) reduces to a scalar Bose-Hubbard Hamiltonian with and effective $U_0$ given by 
$U_0+U_2$.  

Also for $U_{2}>0$, Fig.~\ref{fig-2} shows that the spin interaction leads to a bifurcation, but the explanation is not as easy as in the $U_{2}<0$ case. 
The mechanism at the basis of the localization is the creation of local singlets, promoted by the $U_2$ term, 
which competes with the hopping. 

To understand this mechanism, let us consider the case of an even number
of particles $N=2N_{S}$ with $U_0<0$, in the limit where $U_{2}$ is the dominant energy scale ($U_{2} \gg |U_0|,J$). We also impose that $|U_{0}|  < \frac{4J}{N-1}$ i.e. smaller than the critical point for the bifurcation in the equivalent scalar case (see Appendix~\ref{ap3}).  So the following constraints are satisfied 
$U_{2} \gg  J > (N-1) |U_{0}|/4$. 

In this regime, the hopping can be considered as a perturbation and $U_{0}$ represents the smallest energy
scale. The unperturbed Hamiltonian is:
\begin{equation}
\hat{H}_{2}=\frac{U_{2}}{2}\left(\mathbf{\hat{S}}_{L}^{2}+\mathbf{\hat{S}}_{R}^{2}\right)-U_{2}\hat{N},
\end{equation}
whose ground state is degenerate:
\begin{equation}
 \left|k\right\rangle \equiv \left|s_{L}=0,s_{R}=0,m_{L}=0,n_{L}=2k\right\rangle,
\end{equation}
with $k=0,1,\ldots,N_{S}$. This state represents $k$ singlets in
$L$ and $N_{S}-k$ singlets in $R$. We note that the term 
$\hat{H}_{0}=U_{0}\hat{N}_{L}(\hat{N}_{L}-\hat{N}_{R})+\frac{U_{0}\hat{N}}{2}(\hat{N}-1)$
commutes with $\hat{H}_{2}$ so, even if it is the smallest contribution,
it can be included in the unperturbed Hamiltonian. Moreover, this
term breaks the degeneracy:
\begin{equation}
 \left(\hat{H}_{0}+\hat{H}_{2}\right)\left|k\right\rangle =\epsilon_{0}(k)\left|k\right\rangle \,,
\end{equation}
with
\begin{equation}
 \epsilon_{0}(k)=4U_{0}k\left(k-N_S\right)+U_{0}N_S\left(2N_S-1\right)-2U_{2} N_S.
\end{equation}
The aim is to construct an effective perturbative Hamiltonian in this
subspace:
\begin{equation}
H_{{\rm eff}}=\sum_{k,k'}\epsilon_{k,k'}\left|k\right\rangle \left\langle k'\right|.
\end{equation}
Since the hopping term destroys a singlet, allowing one particle
to move from one site to the other, in order to remain in the singlet subspace the first contribution to the
effective Hamiltonian will be of second order in $J$. Following \cite{cohen}, the form of the effective Hamiltonian is:
\begin{widetext}
\begin{equation}
H_{{\rm eff}}=\sum_{k}\epsilon_{0}(k)\left|k\right\rangle \left\langle k\right|
       - \frac{J^{2}}{2}\sum_{k,k'}\left\langle k\right|\hat{H}_J
       \left[\sum_{\alpha}\left(\frac{1}{\bar{\epsilon}_{0}(\alpha)-\epsilon_{0}(k)}+
       \frac{1}{\bar{\epsilon}_{0}(\alpha)-
       \epsilon_{0}(k')}\right)\left|\psi_{\alpha}\right\rangle 
       \left\langle \psi_{\alpha}\right|\right]\hat{H}_J\left|k'\right\rangle \left|k\right\rangle \left\langle k'\right|,
\end{equation}
\end{widetext}
where
\begin{equation}
\hat{H}_J=\sum_{\sigma=0,\pm1}\left(\hat{a}_{L,\sigma}^{\dagger}\hat{a}_{R,\sigma}+\hat{a}_{R,\sigma}^{\dagger}\hat{a}_{L,\sigma}\right),
\end{equation}
and $\left|\psi_{\alpha}\right\rangle $ are intermediate states with
one singlet broken but still eigenstates of $\left(\hat{H}_{0}+\hat{H}_{2}\right)$
with 
\begin{equation}
\left(\hat{H}_{0}+\hat{H}_{2}\right)\left|\psi_{\alpha}\right\rangle =\bar{\epsilon}_{0}(\alpha)\left|\psi_{\alpha}\right\rangle.
\end{equation}
In our case, $\alpha$ corresponds to a set of indexes $\left\lbrace \sigma,\gamma,k \right\rbrace$ characterizing 
the intermediate states:
\begin{equation}
\left|\psi_{\sigma,\gamma,k}\right\rangle =\left|s_{L}=1,s_{R}=1,m_{L}=\sigma,n_{L}=2k+\gamma \right\rangle, 
\end{equation}
with $\sigma=0,\pm1$ and $\gamma=\pm1$. Note that $\left|\sigma,1,k\right\rangle =\left|\sigma,-1,k+1\right\rangle $ and
$\left|\sigma,-1,k\right\rangle =\left|\sigma,1,k-1\right\rangle$.
So the form of the effective Hamiltonian is:
\begin{eqnarray}
 H_{{\rm eff}}&=&\sum_{k}\Big[D(k)\left|k\right\rangle \left\langle k\right|
\nonumber \\ && + T(k)\big(\left|k-1\right\rangle \left\langle k\right|+\left|k\right\rangle \left\langle k-1\right|\big)\Big] \,
\end{eqnarray}
with 
\begin{eqnarray}
 D(k)&=&\epsilon_{0}(k)-3J^{2}\left(\frac{f^{2}(k)}{\Delta_{k,1}}+\frac{g^{2}(k)}{\Delta_{k,-1}}\right), \\ 
 T(k)&=&-\frac{J^{2}3f(k-1)g(k)}{2}\left(\frac{1}{\Delta_{k,1}}+\frac{1}{\Delta_{k-1,1}}\right),
\end{eqnarray}
\begin{eqnarray}
 f(k) & = & \frac{1}{3}\sqrt{2\left(3N_S+k\left(2N_S-3-2k\right)\right)},\\
 g(k) & = & \frac{1}{3}\sqrt{2k\left(2N_S+3-2k\right)}\\
 \Delta_{k,\gamma} & = & \bar{\epsilon}_{0}(\gamma,k)-\epsilon_{0}(k)=\nonumber \\
                   & = & 2U_{2}+\gamma U_{0}\left[4k+\gamma-2N_S\right].
\end{eqnarray}
This Hamiltonian resembles the scalar one but with the singlets
playing the role of the particle (see Appendix~\ref{ap3}). The hopping term is of the order of
$\frac{T^{2}}{U_{2}}$. It is possible to see numerically that, for
$U_{0}=0$, the ground state energy of $H_{\rm{eff}}$ scales as: 
\begin{equation}
 E_{0}^{\rm{eff}}(U_{0}=0)=-c\frac{J^{2}}{U_{2}}N_{s}^{2} \,,
\end{equation}
where $c$ is a constant of the order of $c\simeq0.7$. The presence
of $U_{0}$ will give a correction 
\begin{equation}
 E_{1}^{\rm{eff}}=U_{0}N_{s}\left(N_{s}-1\right).
\end{equation}
As in the scalar case, $E_{0}^{\rm{eff}}+E_{1}^{\rm{eff}}$ has to be compared
with the atomic limit $\frac{J^{2}}{U_{2}}=0$, giving $\epsilon_{0}(0)=U_{0}N_{s}\left(2N_{s}-1\right)$.
So the condition for the bifurcation is:
\begin{equation}
E_{0}^{\rm{eff}}+E_{1}^{\rm{eff}}\simeq U_{0}N_{s}\left(2N_{s}-1\right),
\end{equation}
which reads:
\begin{equation} \label{eqn:bifcond}
\frac{J^{2}}{U_{2}|U_{0}|}\simeq c.
\end{equation}
It is worth stressing that these expansion is only valid for an even number of particles.
Here, in contrast to what happens in the scalar case, the bifurcation condition seems independent
on the number of particles. This is not completely true, because the condition 
(\ref{eqn:bifcond}) makes sense only if the bifurcation is not reached in the 
corresponding scalar case. This means that, according to (\ref{eqn:bif1}), $|U_{0}| < 4 J/(N-1)$.
and for large $N$ the bifurcation needs higher values of $U_2$ to occur. In the limit of $N \rightarrow \infty$,
there is no distinction between even and odd filling.

\begin{figure*}
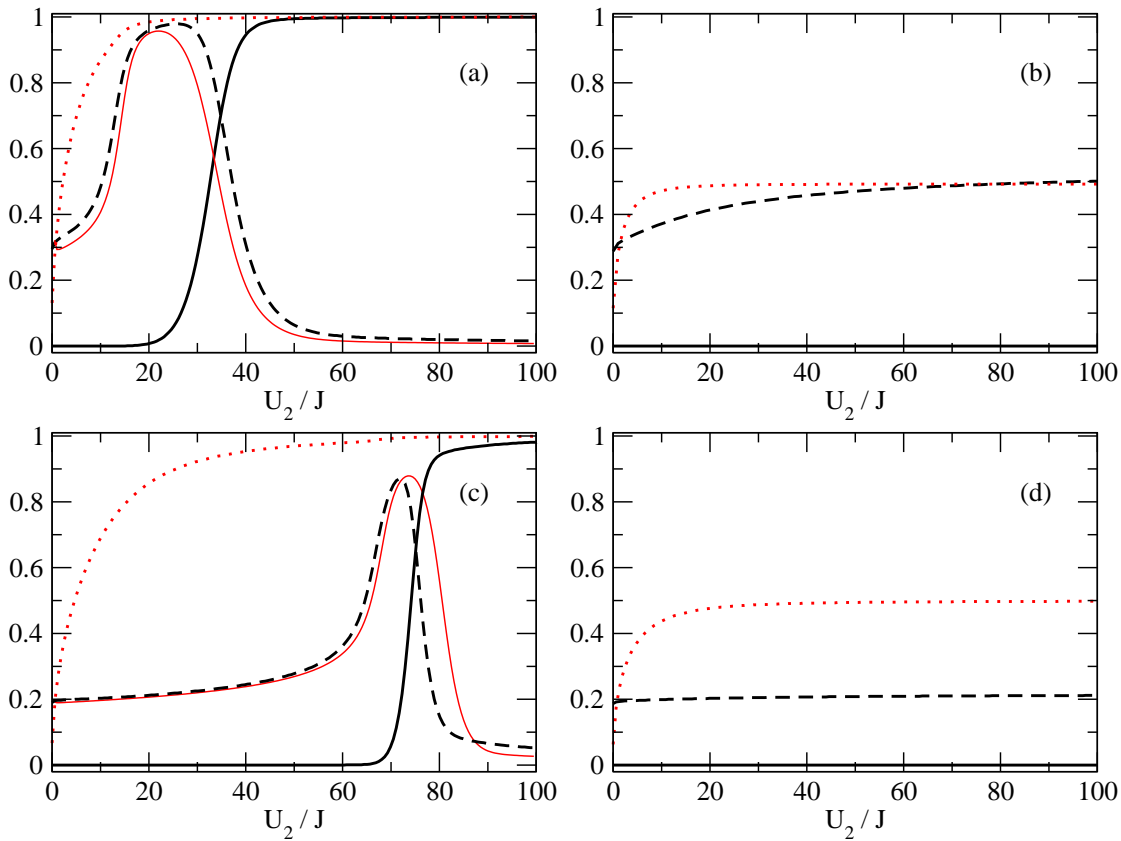

\includegraphics[width=0.85\columnwidth,angle=0, clip=true]{fig4a.eps}
\includegraphics[width=0.85\columnwidth,angle=0, clip=true]{fig4b.eps}
\includegraphics[width=0.85\columnwidth,angle=0, clip=true]{fig4c.eps}
\includegraphics[width=0.85\columnwidth,angle=0, clip=true]{fig4d.eps}
\caption{Expected value of the population imbalance $<\hat{z}>$ (thick-solid) and its dispersion $\sigma_z$ (thick-dashed) as a function of $U_2/J$ and with $U_0/J=-0.05$ for two different number of particles $N=14$ (a), $N=15$ (b), $N=30$ (c) and $N=31$ (d); The analytical dispersion of the population imbalance (thin-solid-red) is also plotted in (a-c), as well as $n_{ps}^L$  (a-c) and $n_{ts}^L$ (b-d) (dotted-red). In all figures we take $\varepsilon_L/J = 10^{-6}$ and $\varepsilon_R/J=0$. }\label{fig-4}
\end{figure*}

\begin{figure*}
\setlength{\unitlength}{0.1\textwidth}
\begin{picture}(9,7.)
\put(0,3.5){
\includegraphics[width=0.85\columnwidth,angle=0, clip=true]{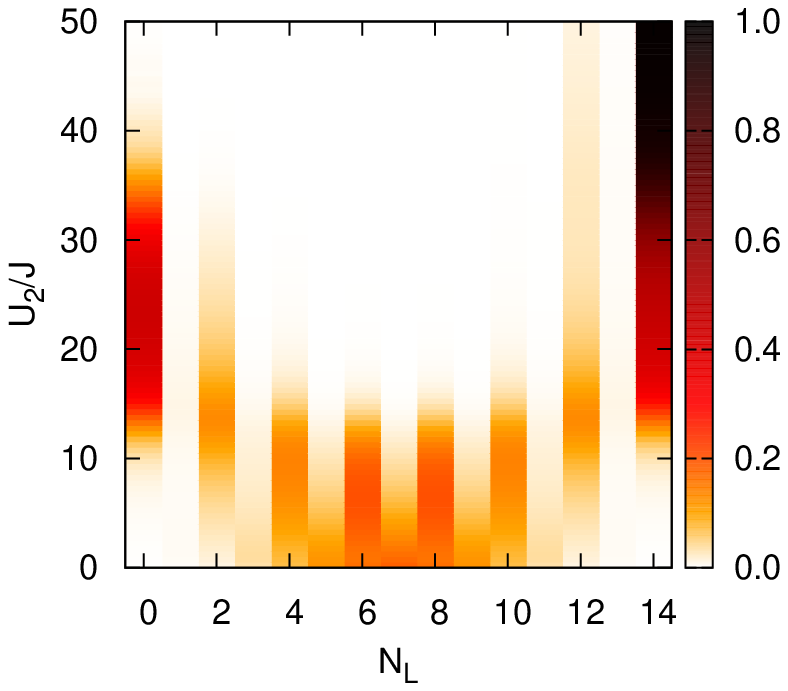}}
\put(1,6.5){\bf{(a)}}
\put(4.5,3.5){
\includegraphics[width=0.85\columnwidth,angle=0, clip=true]{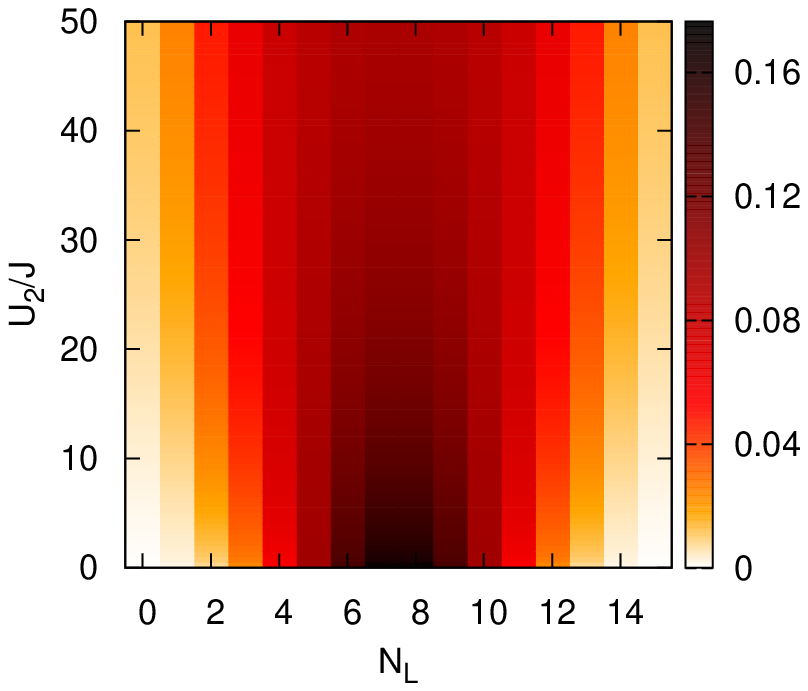}}
\put(5.5,6.5){\bf{(b)}}
\put(0,0){
\includegraphics[width=0.85\columnwidth,angle=0, clip=true]{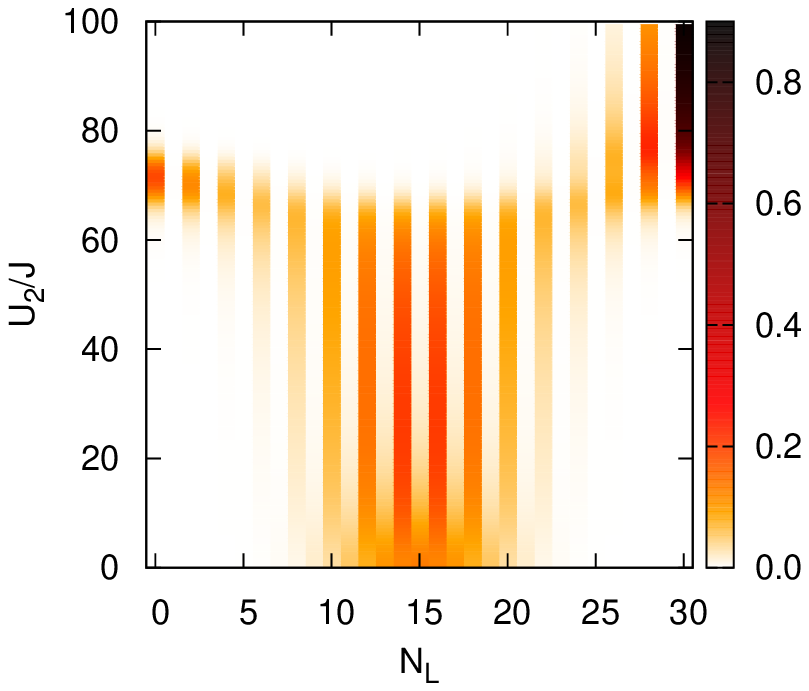}}
\put(1,3){\bf{(c)}}
\put(4.5,0){
\includegraphics[width=0.85\columnwidth,angle=0, clip=true]{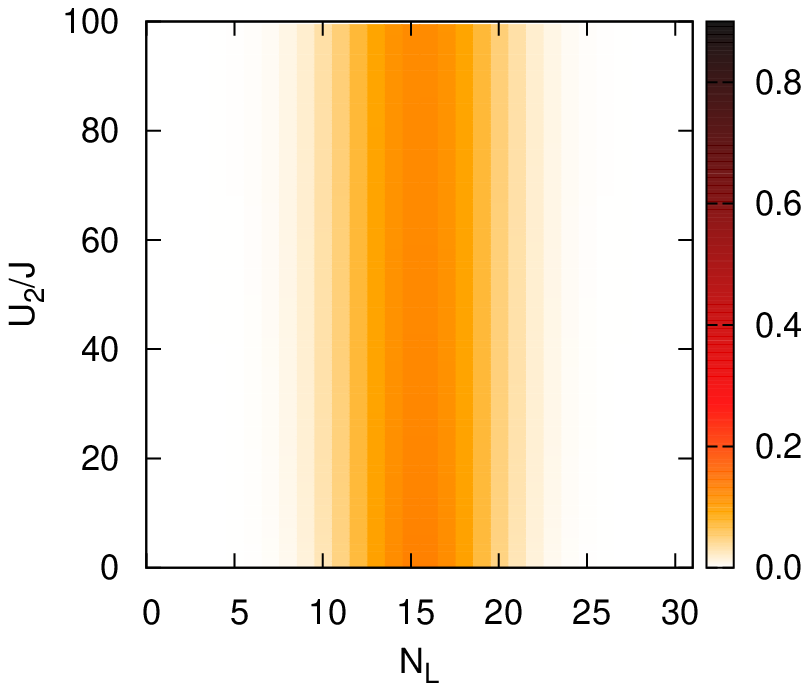}}
\put(5.5,3){\bf{(d)}}
\end{picture}
\caption{Representation of the ground state of a spinor condensate. The color corresponds to the value of the coefficients $|c_{N_L}|^2$, which are plotted as a function of the number of particles in the left well $N_L$ and the interaction $U_2/J$, for $N=14$ (a), $N=15$ (b), $N=30$ (c) and $N=31$ (d). In all figures we take $\varepsilon_L/J = 10^{-6}$ and $\varepsilon_R/J=0$.
}\label{fig-5}
\end{figure*}

The BH model has been studied numerically by  Davidson diagonalization method which allows to find the lowest eigenstates 
of sparse matrices. The diagonalization is carried out in the subspaces with fixed $N$ and $M$.
In all the simulations we take a bias $\varepsilon_L=10^{-6}$ and $\varepsilon_R=0$. 

One can write the ground state isolating the terms including singlets in $L$, $R$ or both.
\begin{eqnarray}
\left|GS\right\rangle &=& \sum_{k}  c_{k}  \left|0,0,2k\right\rangle_L  \left|0,0,N-k\right\rangle_R  +\nonumber \\  
&&  \sum_{k}  d^L_{k}   \left|0,0,2k\right\rangle_L    \left| \nu_k \right\rangle_R   +\nonumber \\ 
&&  \sum_{k}  d^R_{k}  \left| \nu_k \right\rangle_L  \left|0,0,2k\right\rangle_R   + \left| \phi_0  \right\rangle,
\end{eqnarray}
where $\left| \nu_k \right\rangle$ and $\left| \phi_0 \right\rangle$ are not singlet states, 
i.e. they do not have the form (\ref{eqn:singletstate}). 
The component in which both sites are populated only by singlets is referred as 
\emph{pure singlet} component, meaning that it lies in the subspace of singlets defined in the perturbative expansion.  
So, we can define the average density of pure singlets on site $L$ as
\begin{equation}
n^{L}_{ps}=\frac{2}{N}\sum_k k \left| c_k \right|^2,
\end{equation}
and the average density of total singlets on site $L$ as
\begin{equation}
n^L_{ts}=\frac{2}{N}\sum_k k \left( \left| c_k \right|^2+ \left| d_k \right|^2 \right).
\end{equation}
Clearly, if the number of bosons is odd, $n^{L}_{ps}=0$ and singlets can be created only in one site.

Here, we fix a value of $U_0$ corresponding to a state out of the bifurcation in the scalar case, and increase the value of $U_2$.
In Fig.~\ref{fig-4}  (a) and (c) we plot the value of  $n^{L}_{ps}$ for an even number of particle
as a function of $U_2$.  As expected, increasing $U_2$ the density of pure singlets grows and saturates to one, confirming the validity
of the Hilbert space truncation we did in the perturbative expansion.
In the same plots, the population imbalance and its dispersion are also reported, showing the occurrence of 
a quantum analogous to the  bifurcation. As discussed In Sec.~\ref{sec:double}, the bifurcation can be characterized by the inflection point of 
the dispersion, which appears  when almost all the population is constituted by singlets. The exact dispersion is compared with the
one obtained from the effective Hamiltonian, showing a good agreement.  

As previously commented, for any finite number of particles, 
no localization should occur since the spatial symmetry is not broken. Nevertheless, after the bifurcation point, 
the small symmetry breaking induced by the bias is sufficient to localize the condensate. When this occurs, the dispersion of 
the imbalance drops abruptly. 
This phenomenon appears evident looking at the density distribution of the $L$-site (Figs.~\ref{fig-5} (a) and (c)). Here, we observe that the density, 
symmetric and unimodal for small $U_2$,  spreads when increasing $U_2$. At the same time the odd occupation probabilities 
are suppressed because of the population of singlets. At the inflection point, the density becomes flat and starts to be bimodal. 
Then, the bias causes the localization on the left well.

On the other hand, the same analysis can be done for an odd number of particles 
(Figs.~\ref{fig-4} and \ref{fig-5}   (b) and (d)). Here, as stressed before, 
there are no pure singlets components and  the density of total singlet is plotted, showing a saturation to $1/2$. No bifurcation appears 
and the imbalance dispersion does not have inflection points.  This difference between the even and odd cases disappears for a large number 
of particles when no bifurcation should occur, recovering the semiclassical picture where the bifurcation is independent on $U_2$.

\section{Summary and conclusions}

We have studied a spin-1 condensate in a double-well using two-mode approaches of both a mean-field and a fully quantized descriptions. 
First we have presented the mean-field two-mode equations, which conform a system of eight coupled non-linear equations relating the independent variables of the problem. These equations have been used to describe the main features of the ground state, and in future works, will be used to explore in more detail the dynamics of this system. 

Then, starting from the two-site Bose-Hubbard Hamiltonian we have recovered a semiclassical Hamiltonian, and we have found, as expected, that the equations of motion derived from the latter are equivalent to those of the mean-field two-mode approximation. 

We have focused in the study of the ground state properties, and we have found that, for both the mean-field two-mode and the two-site BH descriptions, the number of particles of each component only depends on the sign of the spin-dependent interaction, $U_2$. In each description, the population distribution on each component is equivalent to the corresponding single well distribution, i.e. described by Eqs.~(\ref{eq:mf-u2p}) and (\ref{eq:mf-u2n}) for the mean-field and by Eqs.~(\ref{eq:par1}) and (\ref{eq:par2}) for the fully quantized description. 

Furthermore, we have analyzed the problem of spatial symmetry breaking driven by the spin. For $M=0$, when $U_2<0$ the dependence of the bifurcation with the interactions is well understood and characterized by the mean-field theory. However, when $U_2>0$ the BH model shows a dependence with $U_2$ that the mean-field does not capture. This bifurcation is related to the creation of spin singlets, which drives the symmetry breaking in the system. We have derived an effective Hamiltonian in the double-well potential that describes accurately this transition, and relates it to the total population imbalance and its dispersion.

We acknowledge support from the Spanish MICINN grants  FIS2008-01236 and FIS2008-01661, Generalitat de Catalunya (SGR2009:00343 and 2009-SGR1289), Consolider Ingenio 2010 (CDS2006-00019) and European Regional Development Fund. 
M. M.-M. is supported by an FPI PhD grand of the Ministerio de Ciencia e Innovaci\'on (Spain). S. P. is supported by the Spanish Ministry of Science and Innovation through the program Juan de la Cierva. B. J.-D. is supported by the Ram\'on y Cajal program.

\appendix

\section{Equations for the Two-mode approximation to the time-dependent GP equations}
\label{ap1}

The two-mode equations for a spinor $F=1$ condensate confined in a symmetric double-well potential are a system of eight coupled non-linear differential equations relating the population imbalance and the phase difference of each component $\alpha=0,\pm 1$, defined by
\beqa
z_\alpha(t) &=& {N_{\alpha L}(t)-N_{\alpha  R}(t)\over N_{\alpha}(t)} \\
\delta\phi_{\alpha}(t) &=& \phi_{\alpha R}(t) - \phi_{\alpha L}(t) \;,
\eeqa
the population of the $m=0$ component, $N_0(t)$, and the phase $\Delta\phi_L(t)=2\phi_{0 L}(t) - \phi_{-1 L}(t)-\phi_{+1 L}(t)$. 
The equations obtained, neglecting crossed terms of the left and right modes of the order larger than 1, are  are:
\begin{widetext}
\beqa
\hbar\dot z_{-1} &=&  -2K\sqrt{1-z_{-1}^2} \, \sin{\delta\phi_{-1}} 
+ { \xi_2 N_{0}\sqrt{N_{-1} N_{+1}}\over N_{-1}} \Bigg[
(1-z_{-1})\sqrt{(1+z_{-1})(1+z_{+1})} (1+z_{0})\sin{\Delta\phi_L} \nonumber \\ && -
(1+z_{-1})\sqrt{(1-z_{-1})(1-z_{+1})} (1-z_{0})\sin{\Delta\phi_R}\Bigg]
\nonumber \\ 
\hbar\dot z_{+1} &=&  -2K\sqrt{1-z_{+1}^2} \, \sin{\delta\phi_{+1}} 
+ {\xi_2 N_{0}\sqrt{N_{-1} N_{+1}}\over N_{-1}} \Bigg[
(1-z_{+1})\sqrt{(1+z_{-1})(1+z_{+1})} (1+z_{0})\sin{\Delta\phi_L} \nonumber \\ && -
(1+z_{+1})\sqrt{(1-z_{-1})(1-z_{+1})} (1-z_{0})\sin{\Delta\phi_R}\Bigg]
\nonumber \\ 
\hbar\dot{z}_{0} &=&  -2K \sqrt{1-z_{0}^2} \, \sin{\delta\phi_{0}} 
- 2 \xi_2 \sqrt{N_{-1} N_{+1}} (1-z_{0}^2)\Bigg[
\sqrt{(1+z_{-1})(1+z_{+1})}\sin{\Delta\phi_L}  \nonumber \\ && -
\sqrt{(1-z_{-1})(1-z_{+1})}\sin{\Delta\phi_R}\Bigg] 
\eeqa
\beqa 
\hbar\delta\dot\phi_{-1} &=&
2\xi_0 \sum_\alpha N_\alpha z_\alpha 
+ 2K {z_{-1} \over \sqrt{1-z_{-1}^2}} \cos{\delta\phi_{-1}}
+ 2 \xi_2  \big(N_{-1}z_{-1} + N_{0}z_{0} - N_{+1}z_{+1} \big)
 \nonumber \\
&& - {\xi_2 N_{0}\sqrt{N_{-1}N_{+1}}\over N_{-1} \sqrt{1-z_{-1}^2}} 
\Bigg[\sqrt{(1+z_{-1})(1-z_{+1})}(1-z_{0})\cos{\Delta\phi_R}
-\sqrt{(1-z_{-1})(1+z_{+1})}(1+z_{0}) \cos{\Delta\phi_L}
\Bigg]
\nonumber \\
\hbar\delta\dot\phi_{+1} 
&=& 
2\xi_0 \sum_\alpha N_\alpha z_\alpha 
+2K {z_{+1} \over \sqrt{1-z_{+1}^2}} \cos{\delta\phi_{+1}}
+ 2\xi_2  \Big(- N_{-1}z_{-1}+ N_{0}z_{0}+ N_{+1}z_{+1}\Big)
 \nonumber \\
&& - {\xi_2 N_{0}\sqrt{N_{-1}N_{+1}}\over N_{+1} \sqrt{1-z_{+1}^2}} 
\Bigg[\sqrt{(1-z_{-1})(1+z_{+1})}(1-z_{0})\cos{\Delta\phi_R}
-\sqrt{(1+z_{-1})(1-z_{+1})}(1+z_{0}) \cos{\Delta\phi_L}
\Bigg]
\nonumber \\
\hbar\delta\dot{\phi}_{0}
&=& 
2 \xi_0 \sum_\alpha N_\alpha z_\alpha 
+ 2K {z_{0} \over \sqrt{1-z_{0}^2}}\cos{\delta\phi_{+1}}
+ 2\xi_2 \Big( N_{-1}z_{-1}+  N_{+1}z_{+1}\Big)
 \nonumber \\
&& - 2 \xi_2 \sqrt{N_{-1}N_{+1}} 
\Bigg[\sqrt{(1-z_{-1})(1-z_{+1})}\cos{\Delta\phi_R} 
- \sqrt{(1+z_{-1})(1+z_{+1})}\cos{\Delta\phi_L}\Bigg]
\eeqa
\beqa
\hbar \dot N_{0} &=& - 2 \xi_2 N_{0}\sqrt{N_{-1} N_{+1}} \Bigg[
\sqrt{(1+z_{-1})(1+z_{+1})} (1+z_{0})\sin{\Delta\phi_L}  \nonumber \\ &&
+ \sqrt{(1-z_{-1})(1-z_{+1})} (1-z_{0})\sin{\Delta\phi_R}\Bigg]
\nonumber \\
\hbar\Delta\dot\phi_L
&=&
2 \xi_2 \Big[N_{0} - N_{-1} - N_{+1} + N_{0} z_{0} - N_{-1} z_{-1} - N_{+1} z_{+1} \Big]
\nonumber \\
&& + 2K\sqrt{1-z_{0}\over 1+z_{0}} \cos{\delta\phi_{0}}
 - K\sqrt{1-z_{-1} \over 1+z_{-1}}\cos{\delta\phi_{-1}}
 - K\sqrt{1-z_{+1} \over 1+z_{+1}}\cos{\delta\phi_{+1}}
\nonumber \\
&& - \xi_2 \sqrt{N_{-1}N_{+1}} \sqrt{(1+z_{-1})(1+z_{+1})}\cos{\Delta\phi_L}
\Bigg[4
- {N_{0}(1+z_{0})\over N_{-1}(1+z_{-1})}
- {N_{0}(1+z_{0})\over N_{+1}(1+z_{+1})} \Bigg] \nonumber \\
\eeqa
\end{widetext}
where the parameter
\beqa
K &=&  - \int d^3r \bigg[{\hbar^2\over 2M} \nabla \Phi_L \cdot \nabla \Phi_R
+ \Phi_L V_{\rm ext} \Phi_R \bigg]
\eeqa
takes into account the tunneling between wells, and 
\beqa
\xi_{0(2)} &=& {c_{0(2)}\over 2} \int d^3 r \Phi_{L}^4 (\vec{r})
= {c_{0(2)}\over 2} \int d^3 r \Phi_{R}^4 (\vec{r})
\eeqa
is proportional to the strength of the atom-atom interaction in each well.

Note that there are only 8 independent variables, as the phase $\Delta\phi_R(t)=2\phi_{0 R}(t) - \phi_{-1 R}(t)-\phi_{+1 R}(t)$ can be written as a function of $\Delta\phi_L$ and $\delta\phi_\alpha$, and for a given number of particles $N$ and magnetization $M$, the population of the other components is known $N_{\pm 1}(t) = (N-N_0(t)\mp M)/2$.

These equations reduce to the standard two-mode equations for the scalar case \cite{smerzi} when $N_{0}=N_{-1}=0$, and to the binary mixture without particle interchange when $N_{0}=0$ and $c_2=0$ \cite{smerzi}. 

\section{Semiclassical BH Hamiltonian}
\label{ap2}

The semiclassical approximation to the BH Hamiltonian Eq.~(\ref{eq:bh}) gives the semiclassical Hamiltonian:
\begin{widetext} 
\beqa
H_{\rm s} &=&
- 2J \sum_{\alpha=0,\pm 1} \sqrt{N_{\alpha,L} N_{\alpha_R}}\cos{\delta\phi_\alpha}
+ {U_0\over 2} \big(N_L^2 + N_R^2)
\nonumber \\
&& + {U_2\over 2} \bigg[\big(N_{+1,L}-N_{-1,L}\big)^2 + \big(N_{+1,L}+N_{-1,L}\big)
\big(2N_{0,L}+1\big) + 2N_{0,L} + 4 N_{0,L} \sqrt{N_{+1,L}N_{-1,L}} \cos {\Delta\phi_L}
\nonumber \\ && 
+ \big(N_{+1,R}-N_{-1,R}\big)^2 + \big(N_{+1,R}+N_{-1,R}\big)
\big(2N_{0,R}+1\big) + 2N_{0,R} 
 + 4 N_{0,R} \sqrt{N_{+1,R}N_{-1,R}} \cos {\Delta\phi_R} \bigg] \,.
\eeqa
\end{widetext}

\section{Scalar BH model: estimation of the bifurcation point}
\label{ap3}

The scalar two-site BH model for  $N$ particles is
\begin{equation}
H=-J\left(\hat{a}_{L}^{\dagger}\hat{a}_{R}+\hat{a}_{R}^{\dagger}\hat{a}_{L}\right)+U_{0}\hat{n}_{R}\left(\hat{n}_{R}-\hat{n}\right)+\frac{U_{0}\hat{n}}{2}\left(\hat{n}-1\right),
\end{equation}
where the last term can be considered as a constant since it commutes
with the whole Hamiltonian. The Hilbert space is spanned by the complete
basis
\begin{eqnarray}
\left|n_{R}\right\rangle \left|n_{L}\right\rangle =\left|n_{R}\right\rangle \left|N-n_{R}\right\rangle, 
\end{eqnarray}
which, being labeled by only one quantum number, can be denoted as
\begin{eqnarray}
\left|\nu\right\rangle =\left|n_{R}\right\rangle \left|N-n_{R}\right\rangle ,
\end{eqnarray}
with $\nu=n_{R}=0,1,\ldots,N$. In terms of projectors the Hamiltonian
reads
\begin{eqnarray}
H & = & \sum_{\nu=0}^{N}\Big[T_s(\nu)\left(\left|\nu-1\right\rangle \left\langle \nu\right|+\left|\nu\right\rangle \left\langle \nu-1\right|\right)
\nonumber \\ 
&& + D_S(\nu)\left|\nu\right\rangle \left\langle \nu\right| \Big],
\end{eqnarray}
with
\begin{eqnarray}
D_s(\nu) & = & U_{0}\left[\nu\left(\nu-N\right)+\frac{N}{2}\left(N-1\right)\right], \\
T_S(\nu) & = & -J\sqrt{\nu(N+1-\nu)},
\end{eqnarray}

For $U_{0}<0$ there is a bifurcation point where
the ground state starts to be self trapped. Here we give a raw estimation
of the value of this point. To start with, since the bifurcation occurs
for $|U_{0}|\ll J$, we consider the free particle case with the interaction
to be treated as a perturbation. 

For $U_{0}=0$ the ground state has energy 
\begin{equation}
E_{0}=-JN,
\end{equation}
the interaction can be added perturbatively
\begin{equation}
E_{1} =U_{0}\frac{N(N-1)}{4}.
\end{equation}
In the other limit, $J=0$, the ground state is degenerate 
with energy
\begin{equation}
\epsilon_{0}=\frac{U_{0}N}{2}\left(N-1\right),
\end{equation}
with a zero correction in the first order in t. Bifurcation is expected
to occur when
\begin{equation}
E_{0}+E_{1}\simeq\epsilon_{0}.
\end{equation}
So, the bifurcation condition is given by
\begin{equation}\label{eqn:bif1}
\frac{(N-1)|U_{0}|}{4J}\simeq1.
\end{equation}
For $N \rightarrow \infty$, the bifurcation occurs also for an infinitesimal value of $U_0$.


\begin{thebibliography}{9}

\bibitem{davis} K. B. Davis, M. O. Mewes, M. R. Andrews, N. J. van Druten, D. S. Durfee, D. M. Kurn, and W. Ketterle, Phys. Rev. Lett. 
{\bf 75}, 3969-3973 (1995). 
{\it \mostra{Bose-Einstein condensation in a gas of sodium atoms}} 

\bibitem{anderson} M. H. Anderson, J. R. Ensher, M. R. Matthewa, C. E. Wieman, and E. A. Cornell, Science {\bf 269}, 198-201 (1995). 
{\it \mostra{Observation of Bose-Einstein condensation in a dilute 
atomic vapor}}

\bibitem{bradley} C. C. Bradley, C. A. Sackett, J. J. Tollett, and R. G. Hulet,
Phys. Rev. Lett. {\bf 75}, 1687-1690 (1995).
{\it \mostra{Evidence of Bose-Einstein condensation in an atomic gas with attractive interaction}}

\bibitem{gross} E. P. Gross, Nuovo Cimento {\bf 20}, 454-477 (1961). 
{\it \mostra{Structure of a quantized vortex in boson systems}}

\bibitem{pitaevskii61} L.P. Pitaevskii, Soviet Phys. JETP {\bf 13}, 
451-454 (1961). 
{\it \mostra{Vortex lines in an imperfect Bose gas}}

\bibitem{dalfovo} F. Dalfovo, S. Giorgini, L. P. Pitaevskii, and S. Stringari, Rev. Mod. Phys.  {\bf 71}, 463-512 (1999).
{\it \mostra{Theory of Bose-Einstein condensation in trapped gases}}

\bibitem{stenger} J. Stenger, S. Inouye, D. M. Stamper-Kurn, H.-J. Miesner, A. P. Chikkatur, and W. Ketterle, Nature {\bf 396}, 345–348 (1998).
{\it \mostra{Spin domains in ground state Bose-Einstein condensates}}

\bibitem{stamper-kurn} D. M. Stamper-Kurn, M. R. Andrews, A. P. Chikkatur, S. Inouye, H.-J. Miesner, J. Stenger, and W. Ketterle, Phys. Rev. Lett. {\bf 80}
2027–2030 (1998).
{\it \mostra{ Optical confinement of a Bose-Einstein condensate}}

\bibitem{ho}
T.-L. Ho, Phys. Rev. Lett. {\bf 81}, 742 (1998).
{\it \mostra{Spinor Bose Condensates in optical traps}}

\bibitem{ohmi} T. Ohmi and K. Machida, J. Phys. Soc. Japan {\bf 67}, 1822–1825 (1998).
{\it \mostra{Bose-Einstein condensation with internal degrees of freedom in alkali atom gases }}

\bibitem{zhang} W. Zhang, S. Yi, and L. You, New. Journal of Physics {\bf 5}, 77.1-77.12 (2003). 
{\it \mostra{Mean field ground state of a spin-1 condensate in a magentic field}}

\bibitem{albiez} M. Albiez, R. Gati, J. F\"olling, S. Hunsmann, M. Cristiani, and M. K. Oberthaler, Phys. Rev. Lett. {\bf 95}, 010402 (2005).
{\it \mostra{Direct observation of tunneling and nonlinear self-trapping in a single bosonic Josephson junction}}

\bibitem{steinhauer} S. Levy, E. Lahoud, I. Shomroni, and J. Steinhauer, Nature {\bf 449}, 579 (2007). 

\bibitem{polls10} B. Juli\'a-D\'iaz, D. Dagnino, M. Lewenstein, J. Martorell, and A. Polls, Phys. Rev. A {\bf 81}, 023615 (2010).
{\it \mostra{Macroscopic self-trapping in Bose-Einstein condensates: Analysis of a dynamical quantum phase transition}}

\bibitem{polls10b} B. Juli\'a-D\'iaz, J. Martorell, and A. Polls, Phys. Rev. A {\bf 81}, 063625 (2010).
{\it \mostra{Bose-Einstein condensates on slightly asymmetric double-well potentials}}

\bibitem{fisher} M.P.A. Fisher, P.B. Weichman, G. Grinstein, and D.S.Fisher, Phys. Rev. B {\bf 40}, 546-570 (1989).

\bibitem{demler03} A. Imambekov, M. Lukin, and E. Demler, Phys.Rev.A {\bf 68}, 063602 (2003).

\bibitem{smerzi}
A. Smerzi, S. Fantoni, S. Giovanazzi, and S. R. Shenoy, 
Phys. Rev. Lett. {\bf 79}, 4950 (1997).
{\it \mostra{Quantum Coherent Atomic Tunneling between Two Trapped Bose-Einstein Condensates}}.

\bibitem{milburn} G. J. Milburn, J. Corney, E. M. Wright, and D. F. Walls, Phys. Rev. A {\bf 55}, 4318 (1997). 

\bibitem{raghavan} S. Raghavan, A. Smerzi, S. Fantoni, and S. R.  Shenoy, 
Phys. Rev. A {\bf 59}, 620 (1999). 

\bibitem{ananikian} D. Ananikian and T. Bergeman, Phys. Rev. A {\bf 73}, 013604 (2006). 
{\it \mostra{Gross-Pitaevskii equation for Bose particles in a double-well potential: Two-mode models and beyond}}

\bibitem{ashab} S. Ashab and C. Lobo, Phys. Rev. A {\bf 66}, 013609 (2002).

\bibitem{njp11}
M. Mel\'e-Messeguer, B. Juli\'a-D\'iaz, M. Guilleumas, A. Polls, and A. Sanpera,
New Journal of Physics {\bf 13}, 033012 (2011).
{\it \mostra{ Weakly linked binary mixtures of $F=1$ $^{87}$Rb Bose-Einstein 
condensates}}. 

\bibitem{yi}
S. Yi, \"O. E. M\"ustecapl{\i}o\u{g}lu, C. P. Sun, and L. You, Phys. Rev. A {\bf 66}, 011601 (2002).
{\it \mostra{Single-mode approximation in a spinor-1 atomic condensate}} 

\bibitem{muste05} \"O. E. M\"ustecapl{\i}o\u{g}lu, W. Zhang, and L. You, Phys. Rev. A {\bf 71}, 053616 (2005).

\bibitem{muste07} \"O. E. M\"ustecapl{\i}o\u{g}lu, W. Zhang, and L. You, Phys. Rev. A {\bf 75}, 023605 (2007).

\bibitem{pra09}
B. Juli\'a-D\'iaz, M. Mel\'e-Messeguer, M. Guilleumas, and A. Polls,
Phys. Rev. A {\bf 80}, 043622 (2009). 
{\it \mostra{Spinor Bose-Einstein condensates in a double-well: Population transfer and 
Josephson oscillations}}.

\bibitem{demler}
A. Wagner, C. Bruder, and E. Demler, Phys. Rev. A {\bf 84}, 063636 (2011).
{\it \mostra{Spin-1 atoms in optical superlattices: Single-atom tunneling and 
entanglement}}. 

\bibitem{wu}
Y. Wu, Phys. Rev. A {\bf 54}, 4534 (1996).
{\it \mostra{Simple algebraic method to solve a coupled-channel cavity QED model}}

\bibitem{bigelow}
C. Law, H. Pu, and N. Bigelow, Phys. Rev. Lett. {\bf 81}, 5257-5261 (1998).
{\it \mostra{Quantum spins mixing in Spinor Bose-Einstein condensates}}.

\bibitem{senio} A. de-Shalit and I. Talmi, {\it Nuclear Shell Theory} (Dover Publications, Mineola, New York, 2004). 

\bibitem{cohen} C. Cohen-Tannoudji, J. Dupont-Roc, and G. Grynberg, {\it Atom-photon Interactions: Basic Processes and Applications} (Wiley, New York, 1992). 

\end{thebibliography}
\end{document}